\documentclass[a4paper,11pt]{article}
\usepackage{jinstpub} 
\usepackage{lineno}

\pdfoutput=1
\usepackage[utf8]{inputenc}
\usepackage{float}

\usepackage{newtxtext,newtxmath}
\usepackage{graphicx}
\usepackage{url}
\usepackage{hyperref}
\usepackage{color}
\usepackage{enumitem}
\usepackage{subfigure}
\usepackage[dvipsnames]{xcolor}
\hypersetup{allcolors=[rgb]{0.0 0.0 0.6},linkcolor=[rgb]{0.75 0.05 0.05}}
\usepackage{bm}
\usepackage{epsfig}

\usepackage{slashed}
\usepackage{color}
\usepackage{accents}
\usepackage{url}
\usepackage[dvipsnames]{xcolor}
\usepackage{cancel}
\usepackage[normalem]{ulem} 
\usepackage[colorinlistoftodos]{todonotes}
\usepackage{natbib}
\usepackage{bbold}
\usepackage{xspace}

\newcommand{\um}{$\mu$m\xspace}
\newcommand{\uA}{$\mu$A\xspace}

\begin{document}


\author[a]{Christina Wang,}
\emailAdd{christiw@fnal.gov}
\author[a]{Cristi\'an Pe\~na,} 

\author[a,b]{Si Xie,}

\author[c]{Emanuel Knehr,}
\author[c,d]{Boris Korzh,}
\author[c]{Jamie Luskin}

\author[b]{Sahil Patel,}
\author[c]{Matthew Shaw,}
\author[e,f]{Valentina Vega,}

\affiliation[a]{Fermi National Accelerator Laboratory, Batavia, IL 60510, U.S.A.}
\affiliation[b]{California Institute of Technology, Pasadena, CA 91125, U.S.A.}
\affiliation[c]{NASA Jet Propulsion Laboratory, Pasadena, CA 91011, USA}
\affiliation[d]{University of Geneva, 1205 Geneva, Switzerland}
\affiliation[e]{Departamento de F\'isica y Astronom\'ia, Universidad T\'ecnica Federico Santa Mar\'ia, Valpara\'iso 2390123, Chile}
\affiliation[f]{Centro Cient\'ifico Tecnol\'ogico de Valpara\'iso-CCTVal, Universidad T\'ecnica Federico Santa Mar\'ia, Casilla 110-V, Valpara\'iso, Chile}

\title{Temperature-Dependent Characterization of Large-Area Superconducting Microwire Array with Single-Photon Sensitivity in the Near-Infrared}


\abstract{

Superconducting nanowire single photon detectors (SNSPDs) are a leading detector technology for time-resolved single-photon counting from the ultraviolet to the near-infrared regime. 
The recent advancement in single-photon sensitivity in micrometer-scale superconducting wires opens up promising opportunities to develop large area SNSPDs with applications in low background dark matter detection experiments. 
We present the first detailed temperature-dependent study of a 4-channel $1\times1$~mm$^{2}$ WSi superconducting microwire single photon detector (SMSPD) array, including the internal detection efficiency, dark count rate, and importantly the coincident dark counts across pixels.
The detector shows saturated internal detection efficiency for photon wavelengths ranging from 635~nm to 1650~nm, time jitter of about 160~ps for 1060~nm photons, and a low dark count rate of about $10^{-2}$~Hz.
Additionally, the coincidences of dark count rate across pixels are studied for the first time in detail, where we observed an excess of correlated dark counts, which has important implications for low background dark matter experiments.
The results presented is the first step towards characterizing and developing SMSPD array systems and associated background for low background dark matter detection experiments.

}

\maketitle
{
  \hypersetup{linkcolor=black}
  \tableofcontents
}


\section{Introduction}

Superconducting Nanowire Single Photon Detectors (SNSPDs) are at the forefront of single-photon detection technologies, with wide-ranging applications in optical communications~\cite{Mao:18,PhysRevLett.124.070501}, quantum information science~\cite{PRXQuantum.1.020317,Takesue:15, Shibata:14, PhysRevA.90.043804,Najafi_2015,Weston:16}, and astronomy~\cite{PhysRevA.97.032329,PhysRevLett.123.070504, 10.1117/1.JATIS.7.1.011004}.
SNSPDs have demonstrated remarkable performance, including ultra-low energy threshold below 0.04~eV (or above 29$\mu$m)~\cite{Taylor:23}, low dark counts as low as $10^{-5}$~Hz~\cite{7752769,Shibata:14, Chiles:2021gxk}, and timing resolution on the order of picoseconds~\cite{Korzh:2020, Mueller:24}.
These capabilities make SNSPDs a highly promising detector technology for low background dark matter detection experiments.

Until recently, the application of SNSPDs in high energy physics (HEP) experiments has been limited by their small active areas, typically around 100~$\mu \text{m}^2$, comprised of nanowires with widths of 100~nm.
Recent progress in fabricating thin superconducting films has enabled the development of large area (mm$^2$) single photon detectors using micrometer-scale superconducting wires~\cite{10.1063/5.0150282, 10.1063/5.0044057, 10.1117/1.JATIS.7.1.011004}.
We refer to these devices as superconducting microwire single-photon detectors (SMSPDs) in the rest of the paper.
This advancement in large active area makes SMSPD an ideal photosensor to detect single photons in dark matter detection experiments~\cite{BREAD:2021tpx, Chiles:2021gxk}.

A first study of the largest reported active-area SMSPD array ($1\times1$ mm$^2$) with near-IR sensitivity was reported in Ref~\cite{10.1063/5.0150282}, where saturated detection efficiency was measured for 1060~nm photons at 0.8~K and a dark count rate (DCR) of 0.1~Hz was achieved.
In this paper, we improve upon previous measurements by providing a full, detailed characterization of a 4-pixel $1\times1$ mm$^2$ SMSPD array fabricated with tungsten silicide (WSi) films featuring simultaneous multi-pixel readout for the first time.
The SMSPD array under test also features a per-pixel area twice as large, with an increased fill factor of 40\% compared to the previous 25\%.
We report the first temperature dependent measurements of the key metrics of SMSPD arrays from 0.2~K to 1.2~K, including the internal detection efficiency, time jitter, DCR, and importantly, the correlation of dark counts across pixels.

The SMSPD array under test is described in Section~\ref{sec:snspd}.
The results of the PCR, time jitter, and DCR are presented in Section~\ref{sec:pcr}, ~\ref{sec:time} and \ref{sec:dcr}, respectively.
The coincidences of the DCR across pixels are also studied in detail in Section~\ref{sec:coincidence}.
Finally, the summary is presented in Section~\ref{sec:summary}.

\section{Superconducting microwire single photon detector array}
\label{sec:snspd}

\begin{figure}[ht]
	\centering
	\includegraphics[width=1.00\linewidth]{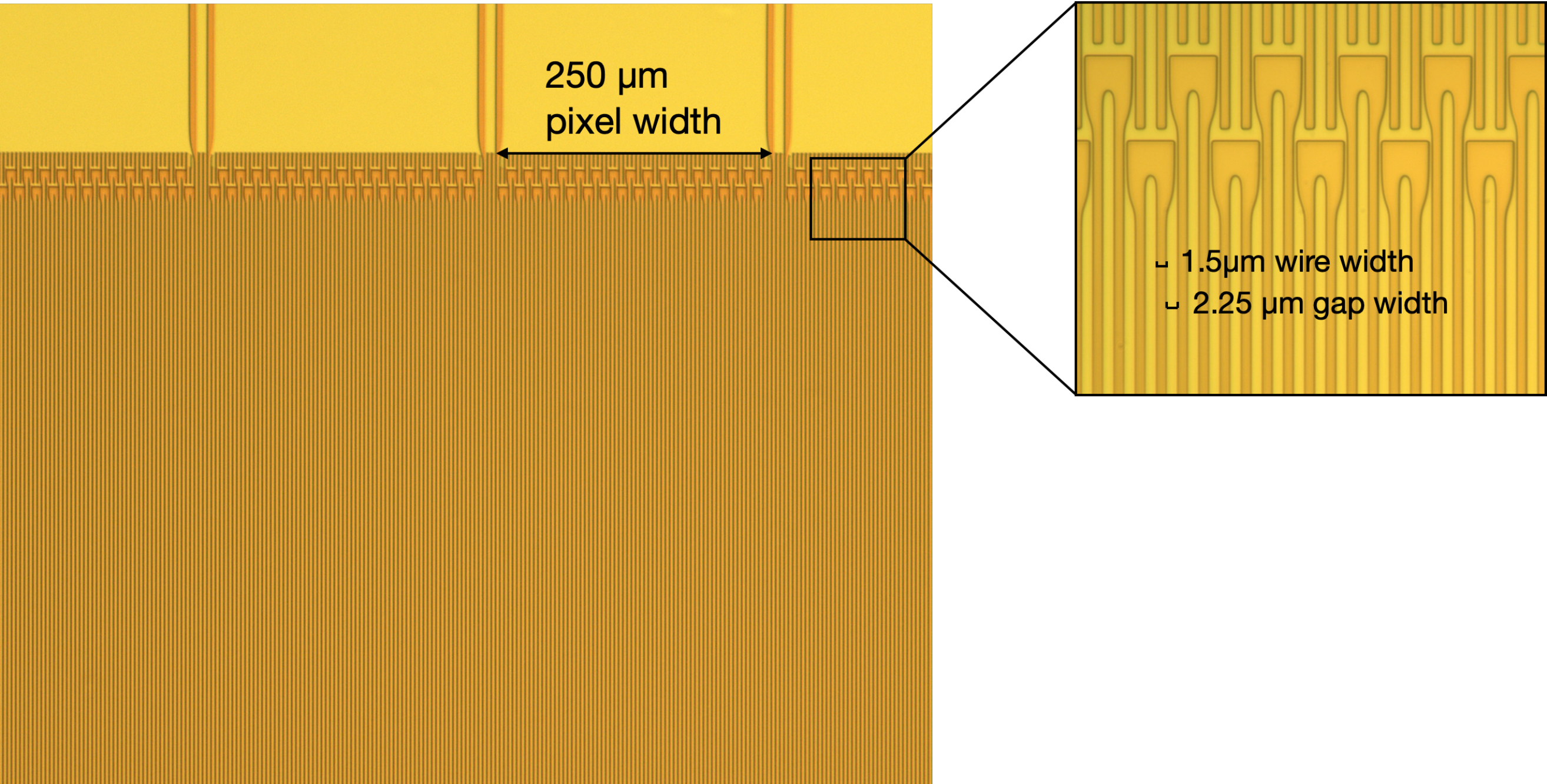}
 	\caption{An optical microscope image of the SMSPD under study is shown.}
	\label{fig:snspd_pic}
\end{figure}

The detector under test is a $1\times1$~mm$^{2}$ 4-channel SMSPD array fabricated on 3~nm thick WSi film. 
The fabrication was carried out at the Jet Propulsion Laboratory.
Each pixel has a size of $0.25\times1$~mm$^{2}$.
The WSi film was sputtered from a $\text{W}_\text{50}\text{Si}_\text{50}$ target and deposited onto an oxidized silicon substrate with a 240~nm-thick oxide.
The thin WSi film was coated with 2~nm of amorphous silicon to prevent oxidation of the WSi film.
The SMSPD were patterned using optical lithography with 1.5~\um-wide wires meandering with a 2.25~\um gap width, amounting to a $40\%$ fill factor.
More details on the optical lithography of SMSPD can be found in Refs.~\cite{10.1063/5.0150282, 6994823}.
Each pixel has an individual single-ended readout.
An optical microscope image of the sensor showing its meandering structure is shown in Figure~\ref{fig:snspd_pic}.

The SMSPD is cooled down to 0.2 -- 1.2~K with an adiabatic demagnetization refrigerator that allows us to tune the operating temperature and has a hold time of about 10-20 hours, depending on the operating temperature.
We biased, amplified, and read out three neighboring pixels of the available four pixels independently. 
A two-stage cryogenic DC-coupled amplifier operating at 40~K was developed by our group to provide a total gain of 30~dB for signal in the 100~MHz to 1~GHz frequency range, similar to that used in Refs.~\cite{10.1063/5.0150282, Pena:2024etu}.
The first stage of the amplifier is based on a low noise high-electron-mobility transistor and the second stage is based on a silicon germanium amplifier.
The DC-coupled amplifier also simultaneously provides SMSPD biasing through the same signal cable.
The signal from each SMSPD pixel is connected through RF cables to room temperature and subsequently recorded by a high-rate time-to-digital converter (Swabian Time Tagger).

Since the bias current of the SMSPD is provided through the DC-coupled amplifier, it not only depends on the bias voltage provided, but also depends on the DC offset of the amplifier and the parasitic resistance of the SMSPD, both of which might vary between different cool downs.
Therefore, the current–voltage characteristic (IV curve) of the SMSPD pixels was measured at the beginning of each cool down to derive the DC offset of the cryogenic amplifier and the parasitic resistance of the SMSPD.

\section{Photon Count Rate Measurement}
\label{sec:pcr}

In this section, the measurement of the internal detection efficiency or the normalized photon count rate (PCR) with external laser excitation is discussed.
To allow a photon from an external laser at room temperature to reach the SMSPD, every layer of radiation shielding has a window with optical filters added to keep the background photon rate low.
Fiber-based laser diodes placed at room temperature are coupled to the cryostat through a reflective collimator that has high reflectivity within the 450~nm--20\um wavelength range and outputs a collimated beam of 8.6~mm in diameter.
The PCR is studied with four fiber-based laser diodes with wavelengths of 635, 1060, 1300, and 1650~nm to understand the SMSPD response to different photon wavelengths.

To be able to accurately measure the PCR and reduce the background count rate from thermal photons, a few optical filters are added at different temperature stages.
A custom cryogenic short-pass filter~\cite{Mueller:21} composed of a N-BK7 glass window with optical coating deposited by Andover Corp., is added to the 40~K stage to filter out photons with wavelengths between 1.9 and 4.5\um with an optical density of 3.
The filter rejects wavelengths shorter than 3\um through the reflective optical coating, and attenuates longer wavelengths through material absorption in the 12.7-mm-thick N-BK7 glass substrate. 
A neutral density filter from Thorlabs is added at the 4~K stage to filter out photons with wavelengths between 1 and 2.6\um with an optical density of 4.
Background rate is reduced to about 1~kHz after the filters are applied.

\begin{figure}[htb!]
	\centering
	\includegraphics[width=0.7\linewidth]{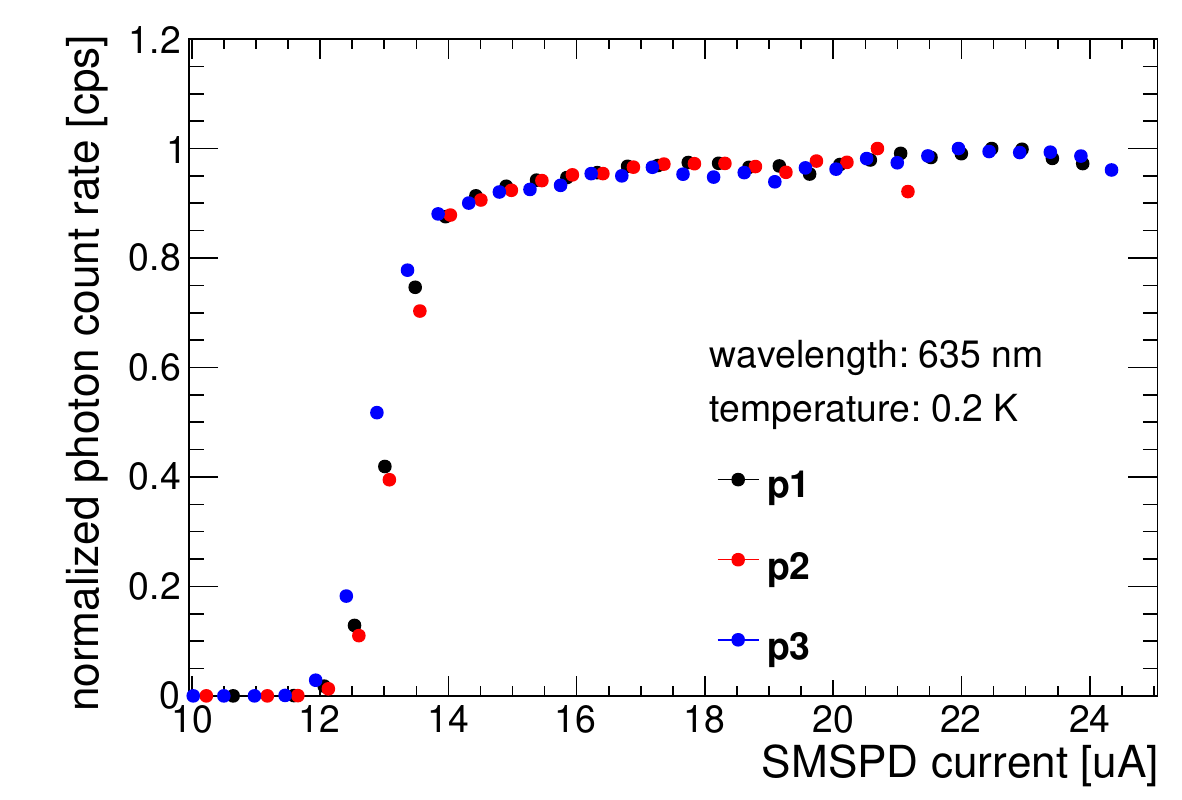}
 	\caption{Normalized PCR measured at 635~nm and 0.2~K for pixel 1, 2, and 3. 
    Similar PCR curves observed for all pixels.
  }
  \label{fig:pcr_pixels}
\end{figure}

The normalized PCR measured at 0.2~K for 635~nm for the three pixels are shown in Figure~\ref{fig:pcr_pixels}.
The normalized PCR is calculated by the PCR subtracted by the background count rate, measured in the same configuration by turning the laser off, and then normalizing the plateau to 1.
The presence of the PCR plateau at high bias current demonstrates that the detector internal detection efficiency is saturated and the bias current above 15$\mu$A is optimal for operation with saturated efficiencies in all three pixels.

\begin{figure}[htb!]
	\centering
	\includegraphics[width=0.45\linewidth]{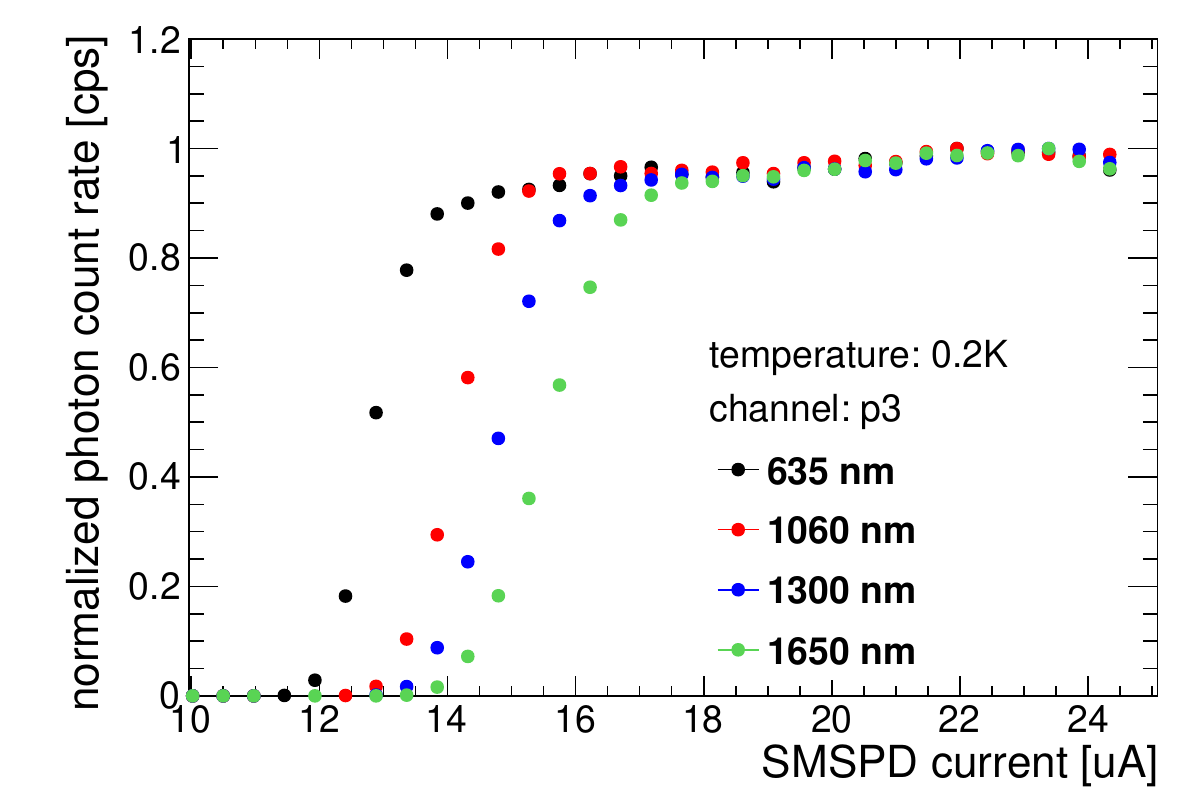}
	\includegraphics[width=0.45\linewidth]{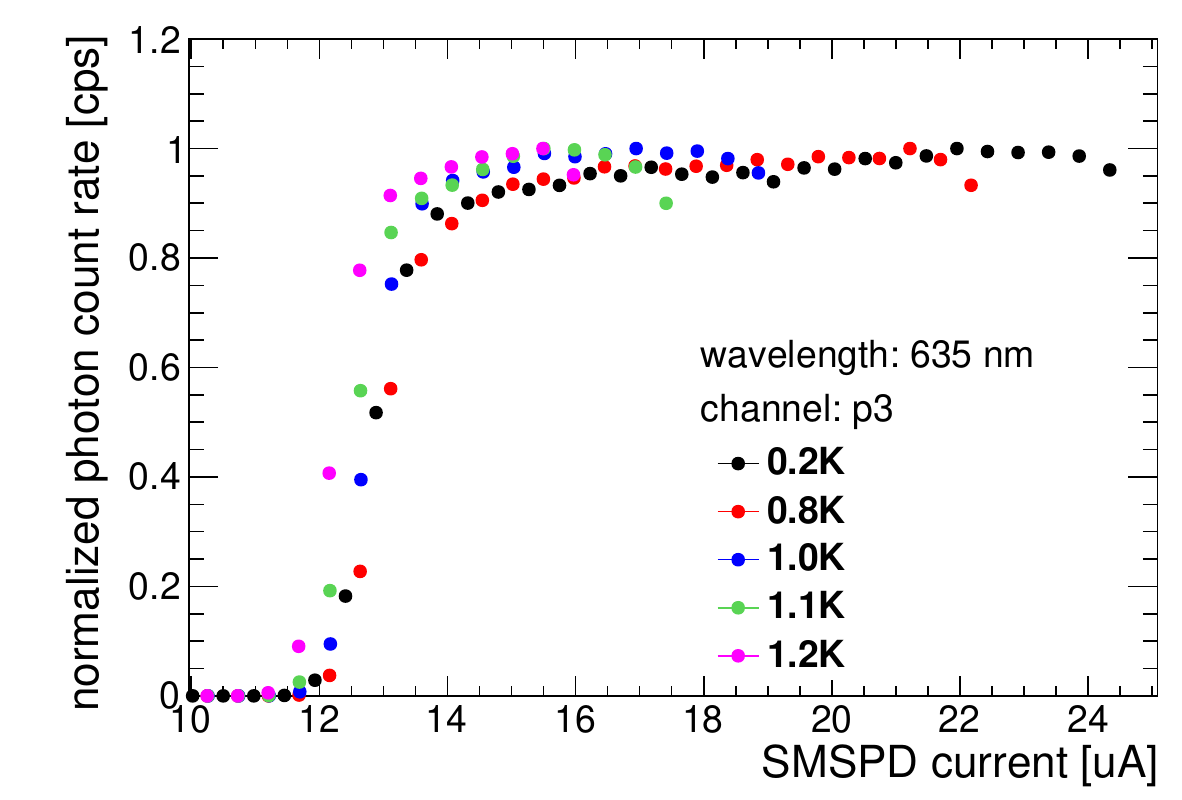}
 	\caption{
    Left: Normalized PCR measured at different wavelengths and at 0.2~K.
    Longer plateau for larger energy as expected.
    Right: Normalized PCR measured at different temperatures and at 635~nm. Similar onset for different temperatures, but longer plateaus for lower energy observed.
  }
  \label{fig:pcr_wl_temp}
\end{figure}

Additionally, the dependence of the normalized PCR with respect to photon wavelength and temperature are also studied. The trend is observed to be similar across all three pixels, and we show the results for pixel 3, the pixel with the longest plateau, as an example in Figure~\ref{fig:pcr_wl_temp}.
As shown in Figure~\ref{fig:pcr_wl_temp} (left), the internal detection efficiency is saturated for all four measured photon wavelengths and the saturation occurs at a lower bias current for higher photon energy.
Figure~\ref{fig:pcr_wl_temp} (right) shows the PCR with respect to the bias current, measured at different operating temperature.
As shown in the plot, the length of the plateau increases as the operating temperature decreases, giving us more leverage on the range of operating bias current.

\section{Time Jitter Measurement}
\label{sec:time}

The time jitter of the SMSPD arrays have also been characterized carefully with varying temperature and bias current.
The time jitter is measured with a picosecond pulsed diode laser with a photon wavelength of 1060~nm.
In this measurement, we keep the free space optical configuration the same as the PCR measurement and replace the continuous wave laser diodes with a picosecond pulsed diode laser with 1060~nm that has a time jitter less than a few picoseconds.
The SMSPD and laser trigger waveforms are acquired using a Keysight UXR0104B oscilloscope.
This oscilloscope features four readout channels with a bandwidth of 10~GHz and a sampling rate of 128~GS/s per channel. 
We record events by triggering on the coincidence of the laser trigger and SMSPD signal per channel.
At least 10,000 coincidence events are recorded for each channel at each bias current and operating temperature.

\begin{figure}[htb!]
	\centering
	\includegraphics[width=0.6\linewidth]{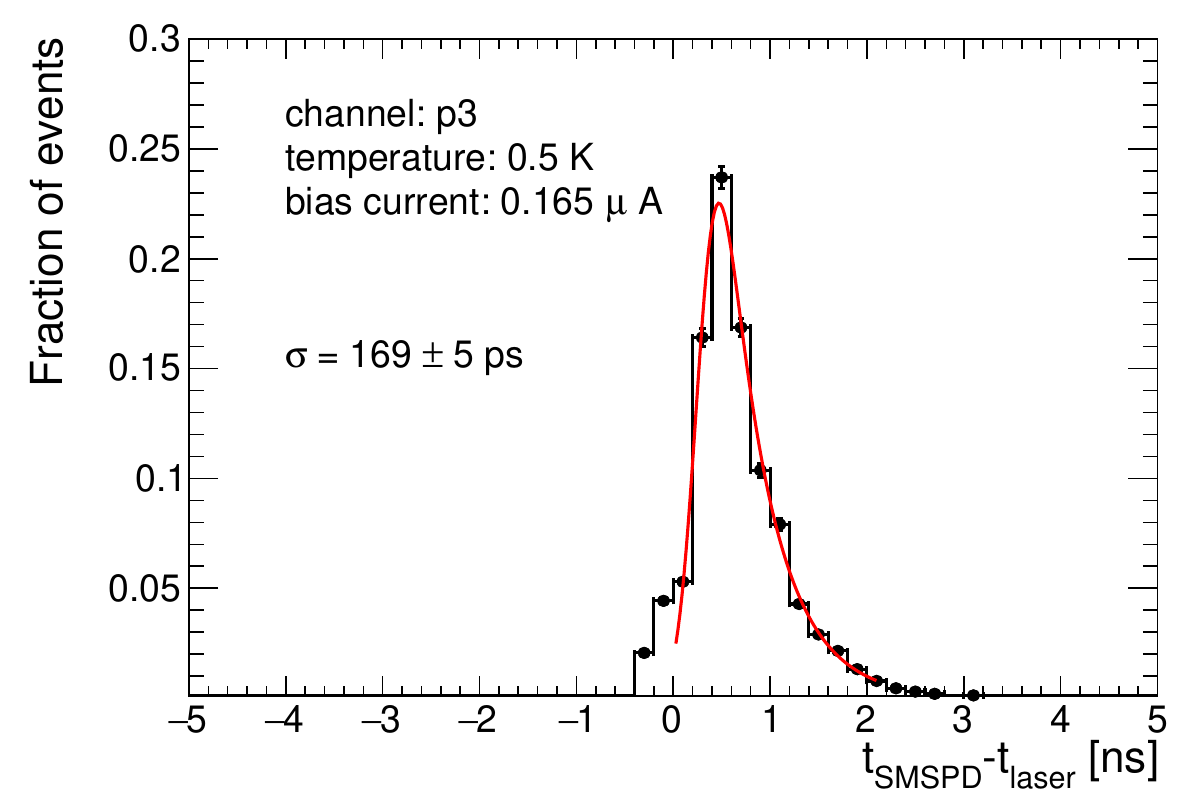}
 	\caption{
    Distribution of time difference between SMSPD signal and laser trigger for p3 measured at an operating temperature of 0.5~K and bias current of 0.165$\mu A$.
    The fitted $\sigma$ of the EMG distribution is $169\pm 5$~ps.
  }
  \label{fig:time_diff}
\end{figure}

The timestamps of the SMSPD and laser trigger are both determined by performing a fit to the rising edge of the pulse to extract the time at which the pulse reaches a fixed threshold, as the SNSPD and laser trigger amplitude are not expected to change.
The fixed threshold is determined separately for each channel, bias current, and operating temperature by calculating half the mean amplitude of all waveforms in the sample.
The time difference between one of the SMSPD pixels and the laser trigger for a particular operating configuration is shown in Figure~\ref{fig:time_diff}.
The time jitter is determined by fitting an exponentially modified Gaussian (EMG), which fits the distribution much more closely than a normal distribution and was also adopted in previous studies~\cite{PhysRevB.96.184504,Caloz:2017gbz,Korzh:2020}.
The fitted $\sigma$ of the EMG distribution shown is $169\pm 5$~ps.
For large area micro-wire arrays, we expect the time jitter to be dominated by geometric jitter caused by time delays introduced by particle detections at different longitudinal positions~\cite{Korzh:2020}.
This result is in agreement with the 138~ps resolution measured for 1064~nm photons in the previous result on SMSPD array~\cite{10.1063/5.0150282} that has a shorter and narrower microwires.
In the future, we plan to improve the time jitter by engineering devices with faster rise times and optimizing the readout scheme to correct for the longitudinal geometric jitter.

\begin{figure}[htb!]
	\centering
	\includegraphics[width=0.45\linewidth]{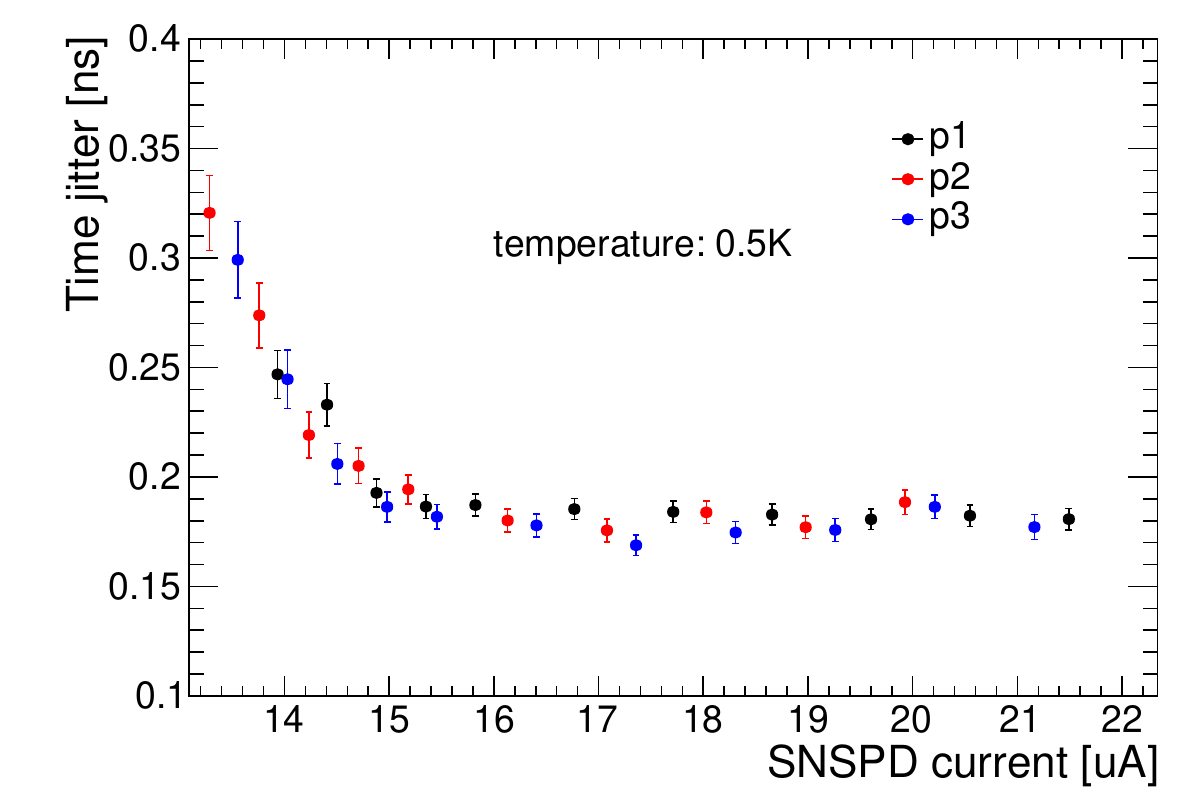}
	\includegraphics[width=0.45\linewidth]{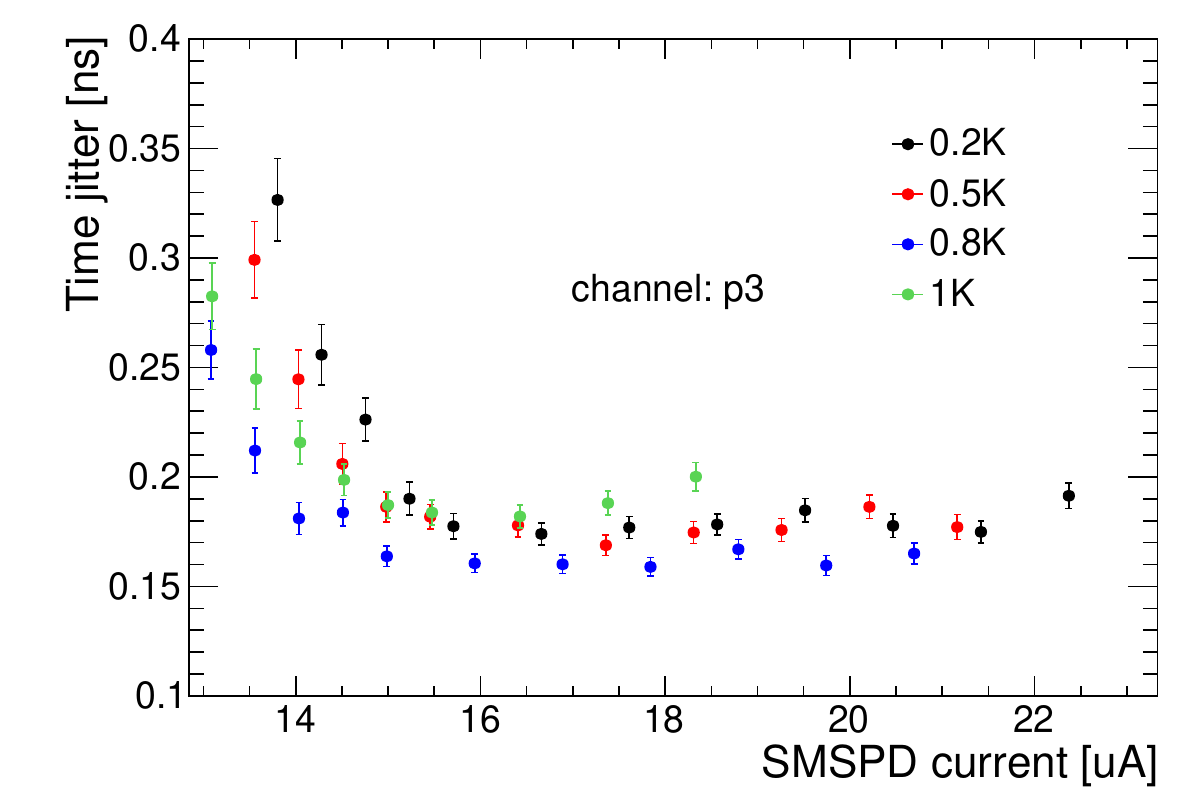}
 	\caption{
    Left: Time jitter with respect to bias current for all three pixels at 0.5~K.
    Right: Time jitter with respect to bias current for pixel 3 measured at different operating temperatures.    
    The time jitter increases and varies at lower bias currents near the onset of detection efficiency, due to rapid changes and variations in the signal-to-noise ratio in that region.
  }
  \label{fig:time_res}
\end{figure}

The time jitter, obtained by the fitted $\sigma$ of EMG, with respect to the bias current comparing different pixels and different operating temperatures are shown in Figure~\ref{fig:time_res}.
The time jitter shows a slight improvement as the internal detection efficiency increases with the bias current and plateaus at higher threshold.
This increasing trend is attributed to the increasing signal-to-noise ratio of the SMSPD waveforms.
This trend is observed to be similar across different SMSPD pixels and different operating temperatures.

\section{Dark Count Rate Measurement}
\label{sec:dcr}

The DCR of the SMSPD is one of the most important characteristics to be measured for a low background dark matter detection experiment.
The DCR is measured by counting the number of SMSPD pulses per second using the Swabian Time Tagger.
To be able to measure the lowest DCR possible, we ensure that the radiation shield at each temperature is light tight to prevent black-body radiation originating from a high temperature stage to propagate towards the detector.
Additionally, a detector lid is added to the SMSPD to prevent background thermal photons from reaching the SMSPDs.

\begin{figure}[htb!]
	\centering
	\includegraphics[width=0.45\linewidth]{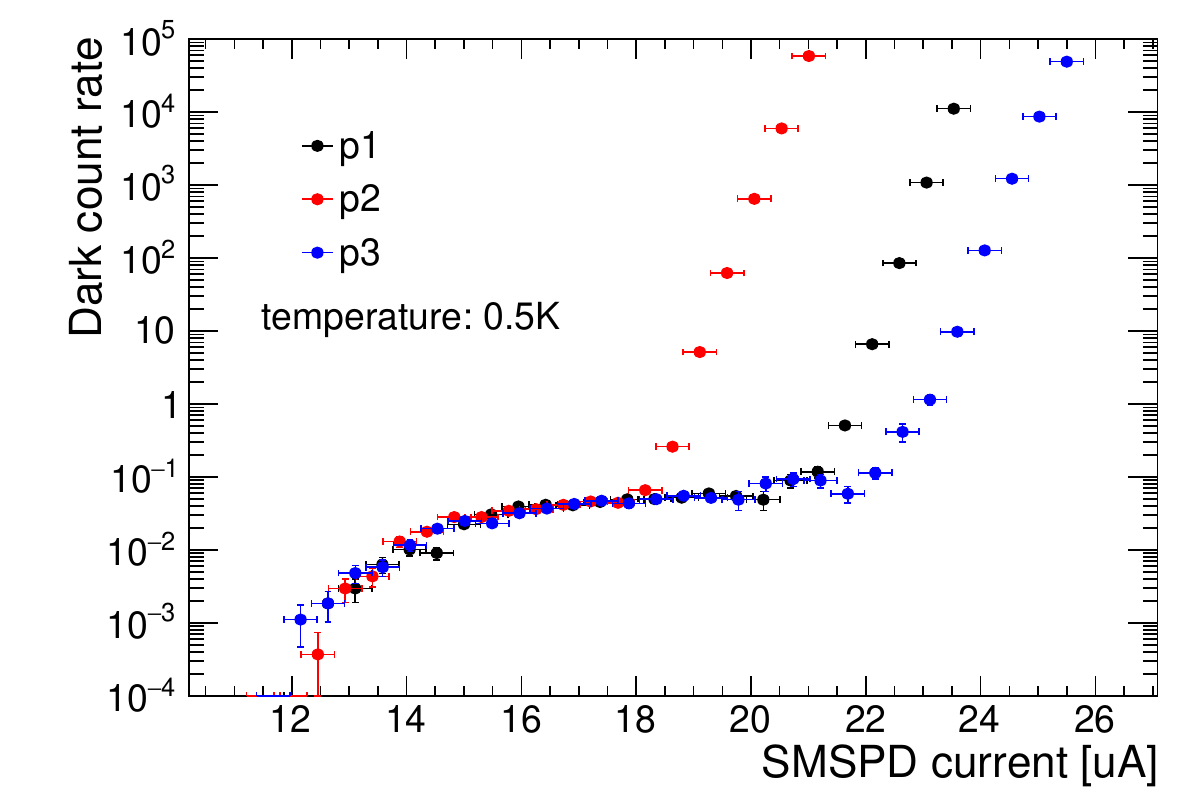}
	\includegraphics[width=0.45\linewidth]{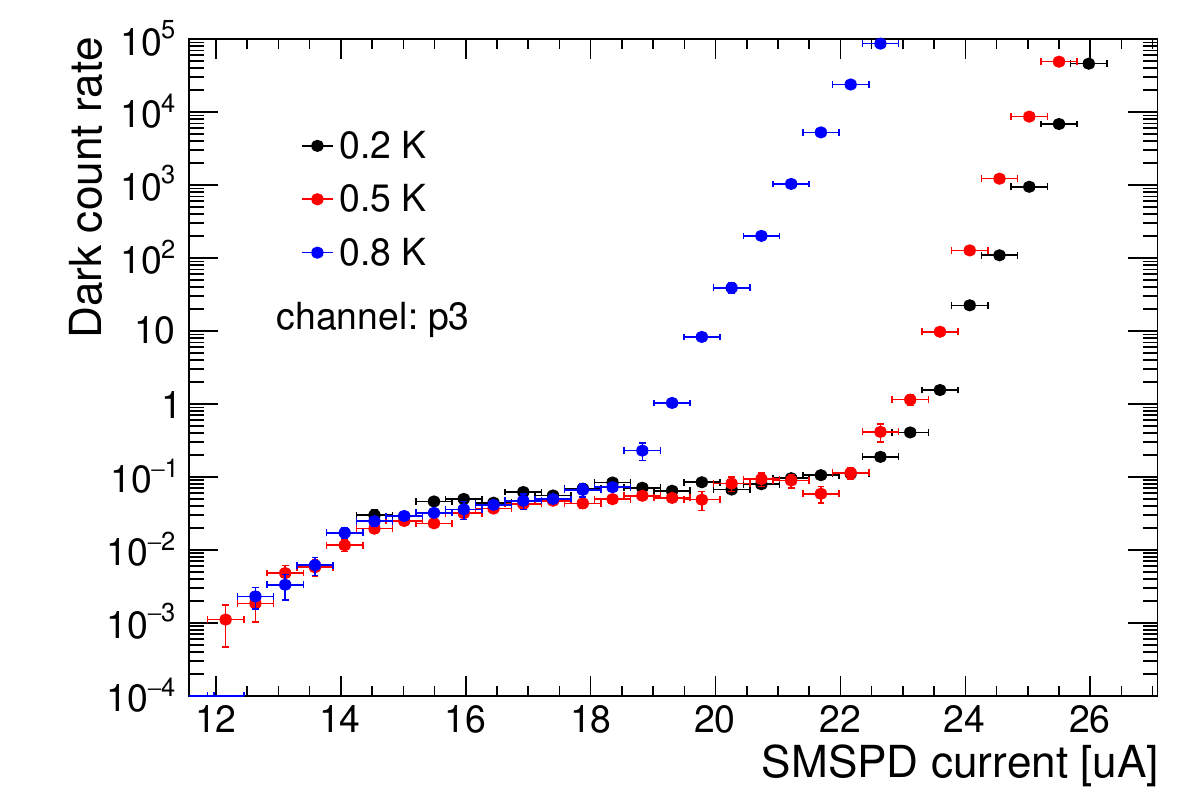}
 	\caption{
    Left: DCR of pixel 1, 2, and 3 at 0.5~K.
    Right: DCR of pixel 3 measured at different operating temperatures.
  }
  \label{fig:dcr}
\end{figure}

The radiation shields can lower the DCR by five orders of magnitude, allowing us to achieve an unprecedentedly low DCR of $10^{-2}$~Hz at 0.2~K, measured for the first time on micron-wide, large-area SMSPDs.
The DCR of all three pixels measured at 0.5~K are shown in Figure~\ref{fig:dcr} (left).
The measured DCR has two distinct regions.
The exponential component at higher bias current corresponds to intrinsic DCR from internal thermal fluctuations.
The origin of the intrinsic dark counts is expected to be current-assisted unbinding of vortex-antivortex pairs~\cite{10.1063/1.3652908}, which has an exponential dependence on the bias current.
 The observed exponential behavior of intrinsic dark counts, is also in agreement with previous studies for narrower wires~\cite{10.1063/1.3652908}.
The flat component at lower bias current approaching $10^{-2}$~Hz suggests that residual background photons are still reaching the SMSPD.

Since the DCR measurements consist of integrating data over a long period of time and across multiple cool downs, the IV curve for the SMSPD pixels were measured at the beginning of every cool down to monitor variations over time.
The parasitic resistance of the SMSPD and the DC offset of the cryogenic amplifier are measured for every cool down and their variation over time results in a systematic uncertainty of 0.28~\uA in the SMSPD bias current, as shown in all figures in this section.

Furthermore, we studied for the first time the temperature dependence of DCR for large-area SMSPDs.
Figure~\ref{fig:dcr} (right) shows the DCR with respect to the bias current, measured at different operating temperature.
As shown in the figure, as the operating temperature decreases, the exponential thermal component of the DCR shifts significantly to the right.
Therefore, given the same bias current, operating at lower temperature significantly decreases the DCR, demonstrating the advantage to operate at as low temperature as possible for DM experiments.
This temperature-dependence that we have observed is in agreement with previous studies for narrower wires~\cite{10.1063/1.3652908}.

\section{Coincidences of Dark Counts Across Pixels}
\label{sec:coincidence}
The coincidences of DCRs are measured and studied for the first time for large area SMSPDs by reading out 3 channels independently and simultaneously.
Given the measured three pixels, there are four possible coincidences: pixel 1 and 2, 1 and 3, 2 and 3, and the three-fold coincidence across all three pixels.
The observed coincidences for the four possible combinations are compared against the expected accidental coincidences to understand if there are any additional correlated sources of background.
The relative time delays between channels are corrected before requiring coincidences.
The expected N-fold accidental coincidence is defined as:
\begin{equation}
\text{expected rate} = N \cdot \text{t}^{(N-1)} \prod_{i=1}^{N} \text{DCR}_i 
\end{equation}
where $t$ is the coincidence window and $\text{DCR}_i$ is the DCR from the i-th pixel.

\begin{figure}[htb!]
	\centering
	\includegraphics[width=0.45\linewidth]{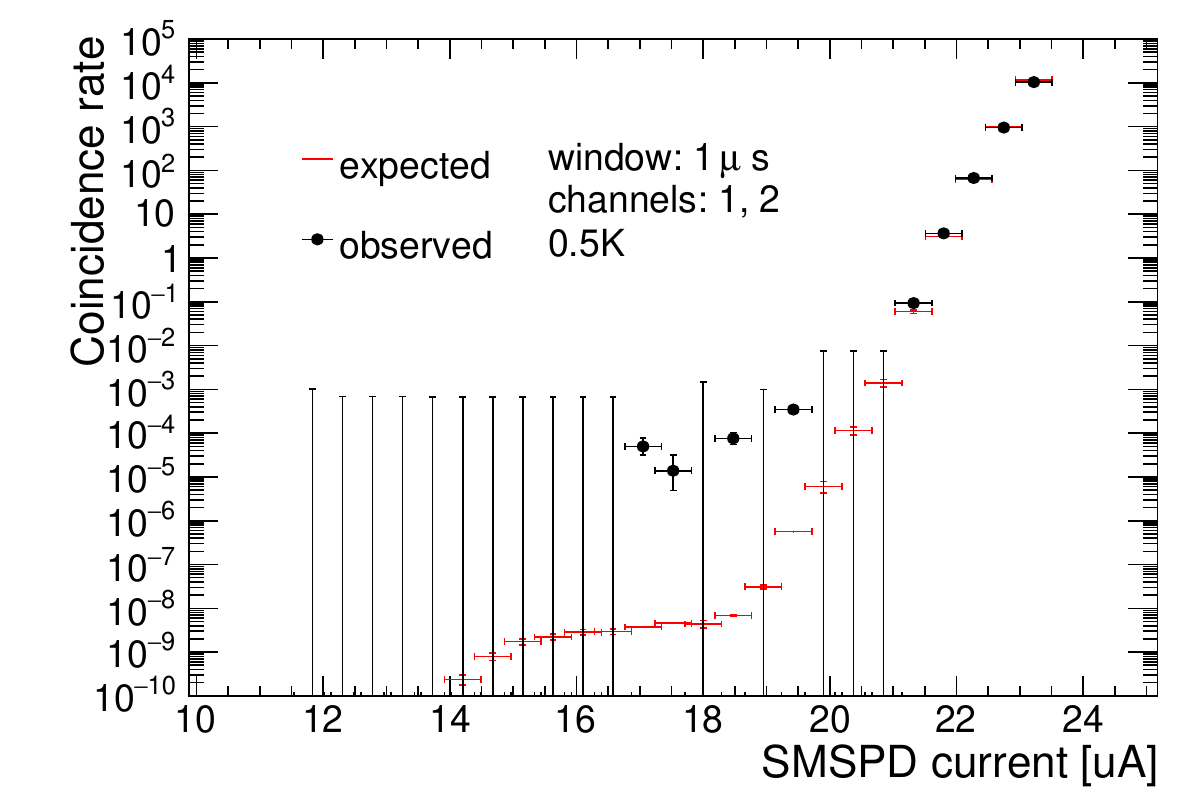}
	\includegraphics[width=0.45\linewidth]{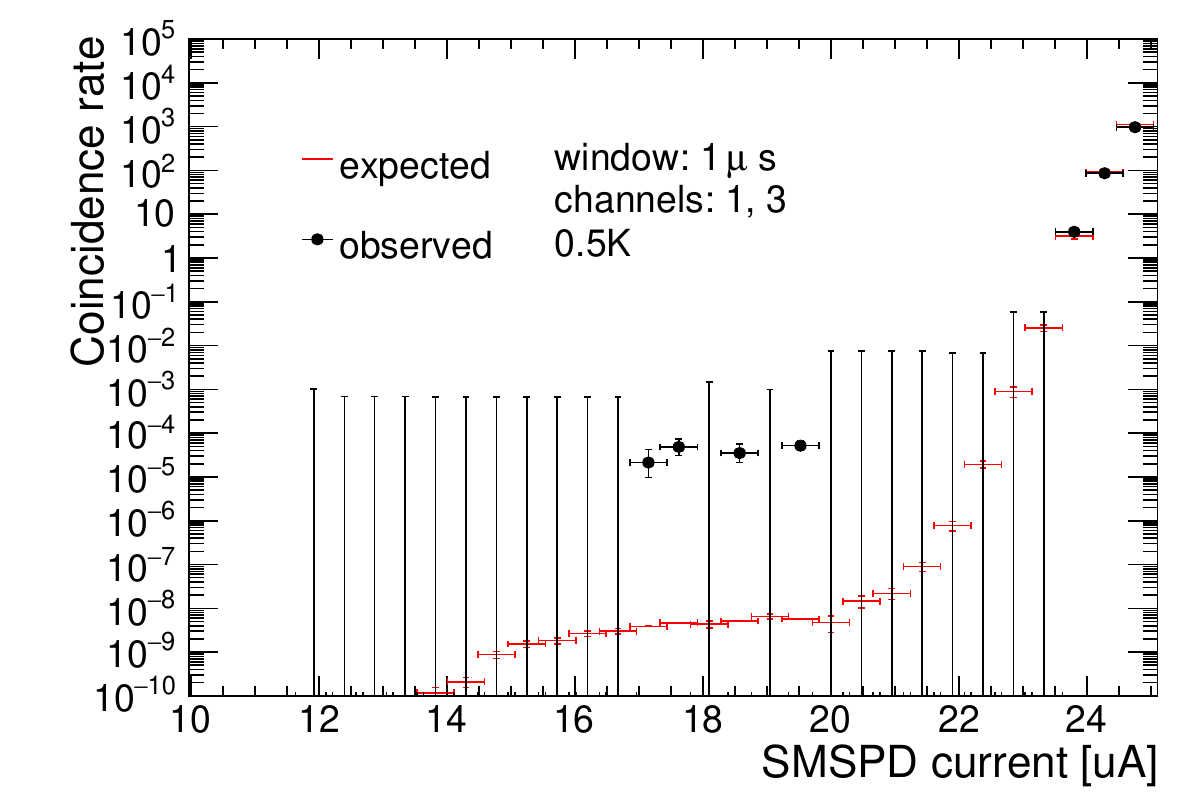}
	\includegraphics[width=0.45\linewidth]{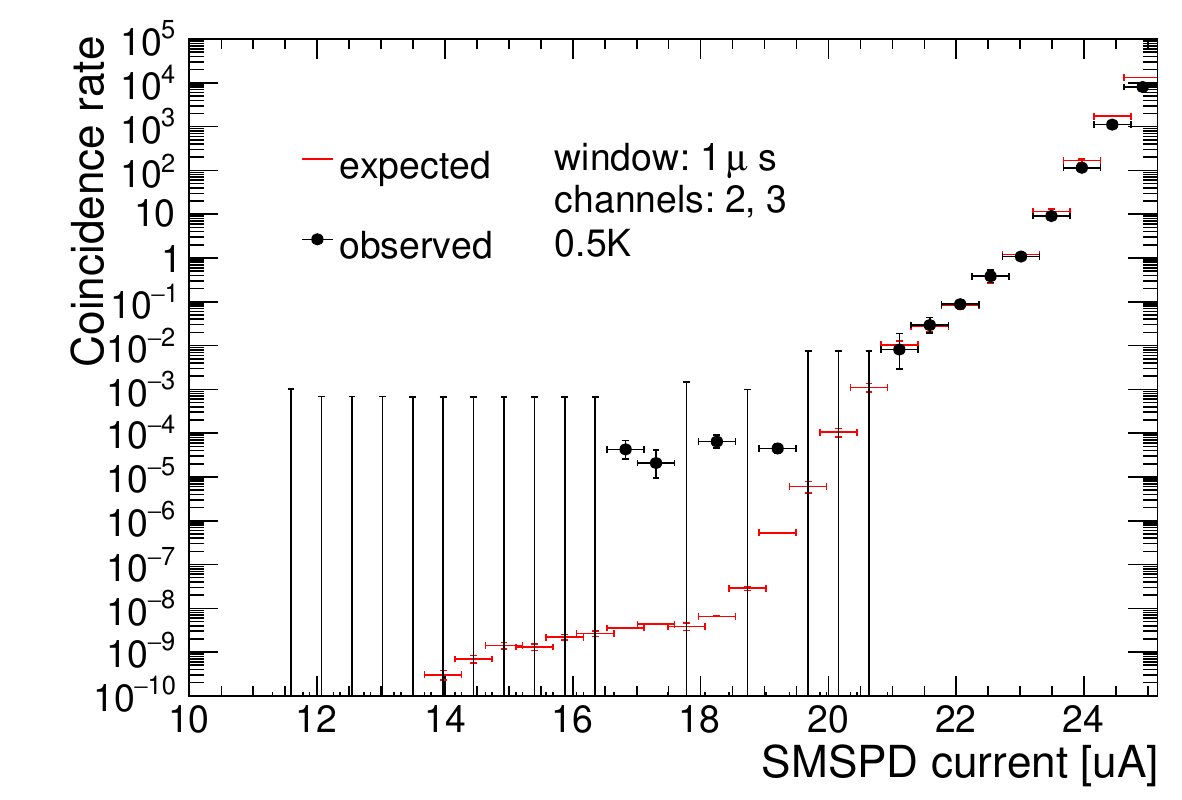}
	\includegraphics[width=0.45\linewidth]{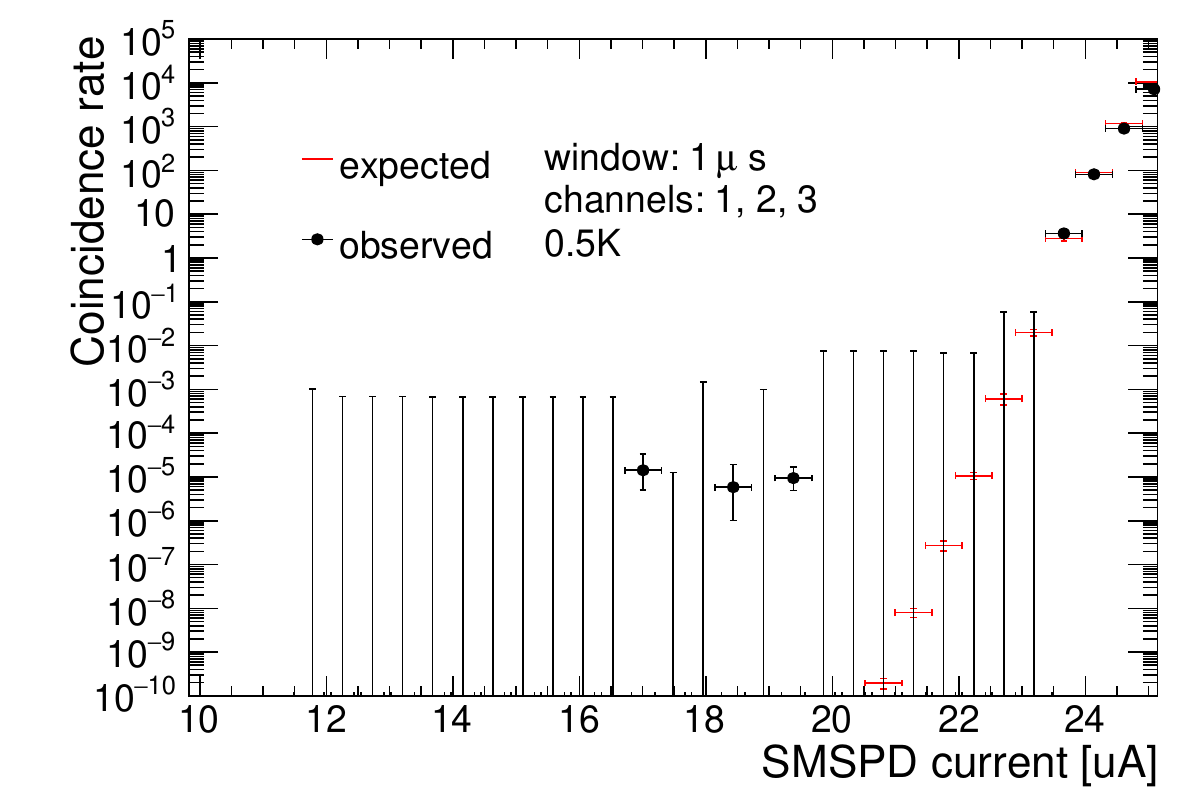}
 	\caption{
    Observed and expected coincidences between pixel 1 and 2 (top left), 1 and 3 (top right), 2 and 3(bottom left), 1, 2, and 3 (bottom right) measured at 0.5K.
    The measurements are performed by varying the SMSPD bias voltages, but the same voltage is applied uniformly across all pixels.
    The x-axis correspond to the average of the SMSPD bias current, which can vary slightly across pixels due to the different parasitic resistance and offset of the DC-coupled amplifier.
    The error bars on the SMSPD bias current correspond go a 0.28$\mu A$ systematic uncertainty attributed to variation of SMSPD bias current across cool downs.
    The error bars on the dark count rate correspond to statistical uncertainties implemented with the Garwood method.
    Measurements with zero observed coincidences were performed with much shorter integration time.
    The large statistical error bars indicate the shorter integration time measurements could not constrain the dark count rate to the observed coincidence level of $10^{-4}$ and $10^{-5}$~Hz.
    Agreement is observed between the observed and expected accidental coincidences for DCR as low as about $10^{-2}$~Hz.
    For bias current below that point, we observe coincidence DCR much higher than the expected rate for all four coincidence combinations.
  }
  \label{fig:coincidence_0p5}
\end{figure}

We first measured the coincidences with a coincidence window of 1~$\mu s$ for varying SMSPD bias current.
The observed and expected coincidences for all four combinations are shown in Figure~\ref{fig:coincidence_0p5}.
The vertical error bars on the dark count rate correspond to statistical uncertainties, which is implemented with the Garwood method~\cite{10.1093/biomet/28.3-4.437} using a 68\% confidence interval which provides coverage for event counts following Poisson distributions, especially when statistics is low.

Agreement is observed between the observed and expected accidental coincidences for DCR as low as about $10^{-2}$~Hz.
However, for bias current below that point, we observe coincidence DCR of about $10^{-4}$ and $10^{-5}$~Hz for the two-fold and three-fold coincidences, respectively, which is much higher than the expected rate for all four coincidence combinations.
The agreement observed at rates above $10^{-2}$~Hz indicates negligible detector crosstalk.
To further rule out crosstalk as a source of the observed coincidence excess, we repeated the background count rate measurement with the same configuration as in Section~\ref{sec:pcr}, where the cryostat is not completely sealed to allow free-space laser coupling.
The results confirmed that crosstalk does not account for the observed signals.

The varying statistical uncertainties across different measurements are due to differences in integration time.
We first perform a short scan of measurement by scanning over the full range of bias current.
Then we perform long integration time measurement at selected bias current where the expected coincidence rate is low.
We typically perform measurement at one bias current throughout the entire 10-12 hour cryostat hold time of the cryostat and then switch to a different bias current in the following cool down.
All measurements with zero observed coincidences shown in the figures were performed with shorter integration time.
The large coverage of these error bars indicates that the integration time at those bias currents were too short to constrain the dark count rate to the observed coincidence level of $10^{-4}$ and $10^{-5}$~Hz.

\begin{figure}[htb!]
	\centering
	\includegraphics[width=0.45\linewidth]{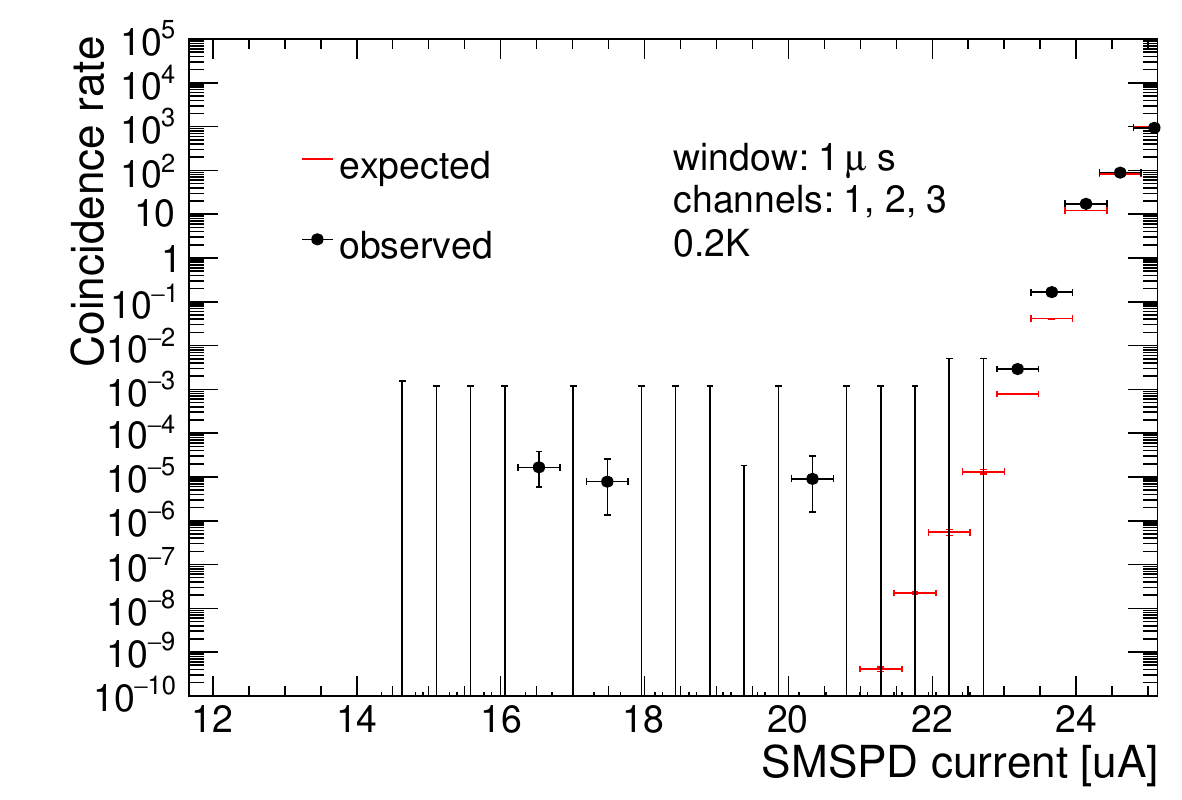}
	\includegraphics[width=0.45\linewidth]{figures/DCR_p1p2p3_0p5K.pdf}
	\includegraphics[width=0.45\linewidth]{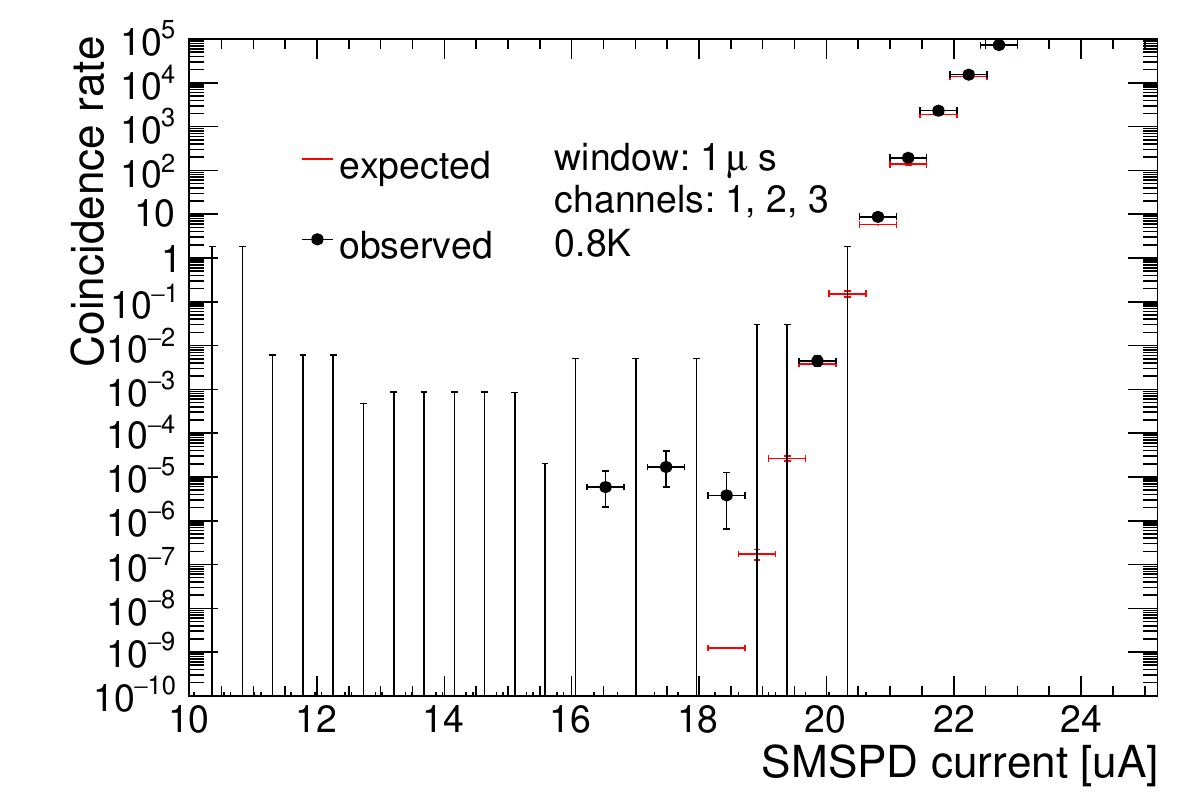}
 	\caption{
    Observed and expected coincidences between pixel 1, 2, and 3 measured at temperature 0.2 (top left), 0.5 (top right), 0.8~K (bottom).
    The measurements are performed by varying the SMSPD bias voltages, but the same voltage is applied uniformly across all pixels.
    The x-axis correspond to the average of the SMSPD bias current, which can vary slightly across pixels due to the different parasitic resistance and offset of the DC-coupled amplifier.
    The error bars on the SMSPD bias current correspond go a 0.28$\mu A$ systematic uncertainty attributed to variation of SMSPD bias current across cool downs.
    The error bars on the dark count rate correspond to statistical uncertainties implemented with the Garwood method.
    Measurements with zero observed coincidences were performed with much shorter integration time.
    The large statistical error bars indicate the shorter integration time measurements could not constrain the dark count rate to the observed coincidence level of $10^{-4}$ and $10^{-5}$~Hz
    For low bias current, similar excess of observed coincidence DCR observed at all operating temperatures.
    }
  \label{fig:coincidence_3fold}
\end{figure}

\begin{figure}[htb!]
	\centering
	\includegraphics[width=0.45\linewidth]{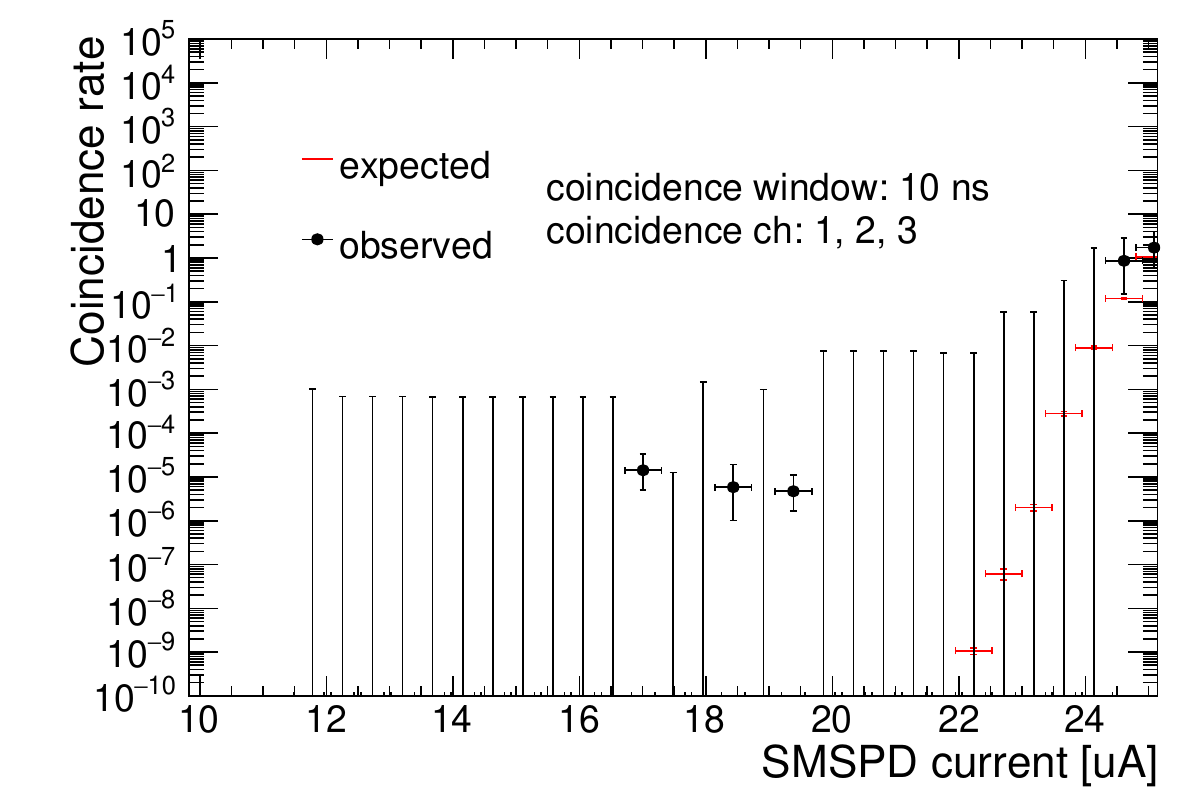}
	\includegraphics[width=0.45\linewidth]{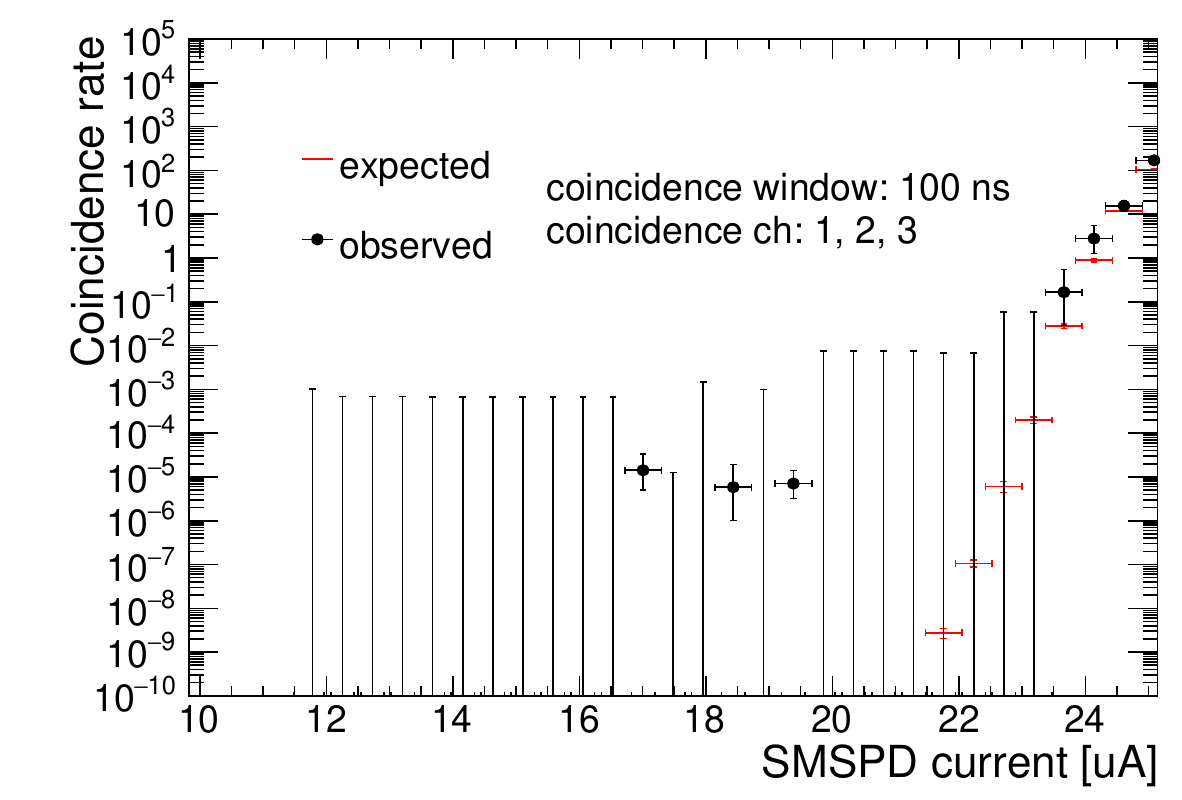}
 	\caption{
Observed and expected coincidences between pixel 1, 2, and 3 measured at temperature 0.5~K for coincidence window of 10~ns (left) and 100~ns (right).
The measurements are performed by varying the SMSPD bias voltages, but the same voltage is applied uniformly across all pixels.
    The x-axis correspond to the average of the SMSPD bias current, which can vary slightly across pixels due to the different parasitic resistance and offset of the DC-coupled amplifier.
    The error bars on the SMSPD bias current correspond go a 0.28$\mu A$ systematic uncertainty attributed to variation of SMSPD bias current across cool downs.
    The error bars on the dark count rate correspond to statistical uncertainties implemented with the Garwood method.
    Measurements with zero observed coincidences were performed with much shorter integration time.
    The large statistical error bars indicate the shorter integration time measurements could not constrain the dark count rate to the observed coincidence level of $10^{-4}$ and $10^{-5}$~Hz.
    The observed coincidence rate doesn't decrease significantly as the coincidence window decreases, suggesting the correlated signals in the different pixels come from a single physical source.
    }
  \label{fig:coincidence_cw}
\end{figure}

Similar trends are also observed at different temperatures, where the three-fold coincidences measured at 0.2, 0.5, and 0.8~K are shown in Figure~\ref{fig:coincidence_3fold}.
Furthermore, we also adjust the coincidence window from the default 1$\mu s$ to 10 and 100~ns to further understand the temporal nature of the coincidence events, as shown in Figure~\ref{fig:coincidence_cw}.
The observed coincidence rate does not decrease significantly as the coincidence window decreases, demonstrating that the time differences across the channels are less than a few ns.
This suggests that the correlated signals in the different pixels come from a single physical source.

One possible explanation for the source is cosmic muon background, which is expected to have a rate of $10^{-4}$~Hz on a mm$^2$ device. 
For the next steps of our studies, we are building a cosmic muon tagging detector with scintillator planes above and below the cryostat to confirm cosmic muon detection in coincidence with the dark counts from the SMSPD array, to shed more light on the source of these correlated events. 
We defer this study to a future paper.

\clearpage

\section{Summary}
\label{sec:summary}

This paper presents a comprehensive temperature-dependent study of a 4-channel, $1\times1$ mm$^2$ WSi SMSPD array. 
The study includes an analysis of the internal detection efficiency, time jitter, DCR, and DCR coincidences across pixels. 
The SMSPD array shows saturated internal detection efficiency for photon wavelengths ranging from 635~nm to 1650~nm, with a low DCR of approximately $10^{-2}$~Hz. 
For the first time, we investigated the DCR coincidences across pixels, revealing an excess of correlated background noise, which may have significant implications for low-background dark matter experiments. 
This result lays the groundwork for characterizing and developing SMSPD array systems for future dark matter detection experiments.

\acknowledgments

This work has been performed at Fermi National Accelerator Laboratory, which is managed by FermiForward Discovery Group, LLC under Contract No. 89243024CSC000002 with the U.S. Department of Energy, Office of Science, Office of High Energy Physics.
This work is supported by the U.S. Department of Energy, Office of Science Accelerate Initiative Program Award under contract FWP FNAL 23-30, and partially supported by the U.S. Department of Energy, Advanced Scientific Computing Research Program Award under contract FWP FNAL 23-24.
Part of this research was performed at the Jet Propulsion Laboratory, California Institute of Technology, under contract with the National Aeronautics and Space Administration.

\bibliography{refs}
\bibliographystyle{jhep}
\end{document}